\pdfoutput=1
\documentclass[aps,prl,reprint,superscriptaddress,longbibliography,hidelinks]{revtex4-1}

\usepackage{graphicx}
\usepackage{amsmath}
\usepackage{hyperref}

\newcommand{\beginsupplement}{%
        \setcounter{table}{0}
        \renewcommand{\thetable}{S\arabic{table}}%
        \setcounter{figure}{0}
        \renewcommand{\thefigure}{S\arabic{figure}}%
}

\begin{document}

\title{Observation of Optomechanical Quantum Correlations at Room Temperature}

\author{T. P. Purdy}
\email[]{thomas.purdy@nist.gov}
\affiliation{Joint Quantum Institute, National Institute of Standards and Technology, Gaithersburg, MD 20899 USA}

\author{K. E. Grutter}
\author{K. Srinivasan}
\affiliation{Center for Nanoscale Science and Technology, National Institute of Standards and Technology, Gaithersburg, MD 20899 USA}

\author{J. M. Taylor}
\affiliation{Joint Quantum Institute, National Institute of Standards and Technology, Gaithersburg, MD 20899 USA}
\affiliation{Joint Center for Quantum Information and Computer Science, University of Maryland, College Park, Maryland 20742, USA}

\date{May 18, 2016}

\begin{abstract}
  By shining laser light through a nanomechanical beam, we measure the beam's thermally driven vibrations and perturb its motion with optical forces at a level dictated by the Heisenberg measurement-disturbance uncertainty relation.  Such quantum backaction is typically difficult to observe at room temperature where the motion driven by optical quantum intensity fluctuations is many orders of magnitude smaller than the thermal motion.  We demonstrate a cross-correlation technique to distinguish optically driven motion from thermally driven motion, observing this quantum backaction signature up to room temperature.  While it is often difficult to absolutely calibrate optical detection, we use the scale of the quantum correlations, which is determined by fundamental constants, to gauge the size of thermal motion, demonstrating a path towards absolute thermometry with quantum mechanically calibrated ticks.
\end{abstract}


\maketitle


Quantum effects in macroscopic mechanical systems are typically difficult to observe under ambient conditions. At room temperature and atmospheric pressure such effects are swamped by thermal and other noise sources.  However, measuring the system must still imprint at least a small, tell-tale quantum signature, exemplified in its simplest form by Heisenberg's microscope thought experiment, where optical probing necessarily perturbs the motion of the object being probed~\cite{Heisenberg1930,Clerk2010}.  Focusing on harmonic oscillators, a variety of quantum backaction effects have been observed for nanomechanical systems in cryogenic environments~\cite{SafaviNaeini2012,Purdy2013,SafaviNaeini2013,Purdy2013b,Weinstein2014,Lecocq2014,Wilson2015,Purdy2015,Meenehan2015,Underwood2015,Teufel2016,Riedinger2016,Sidhar2016} and cold atom optomechanical systems~\cite{StamperKurn2014}.  However, measurements under ambient conditions have remained an experimental challenge.  Here, we describe a method to tease out the small quantum correlations that are produced when optical-measurement-induced motion is in turn written back onto the light probing a nanomechanical resonator.  Despite the several-orders-of-magnitude-smaller scale of these correlations compared to the thermal and measurement noise, we clearly observe this distinctly quantum signature all the way up to room temperature.  By comparing the magnitude of thermally induced mechanical vibrations, which scales in proportion to the temperature, to the size of the quantum correlation, we demonstrate a wide-range, chip-integrated Brownian motion thermometer whose scale is referenced to fundamental constants, providing a path toward a quantum primary temperature standard.

\begin{figure}
	\begin{center}
	\includegraphics[width=3.1 in]{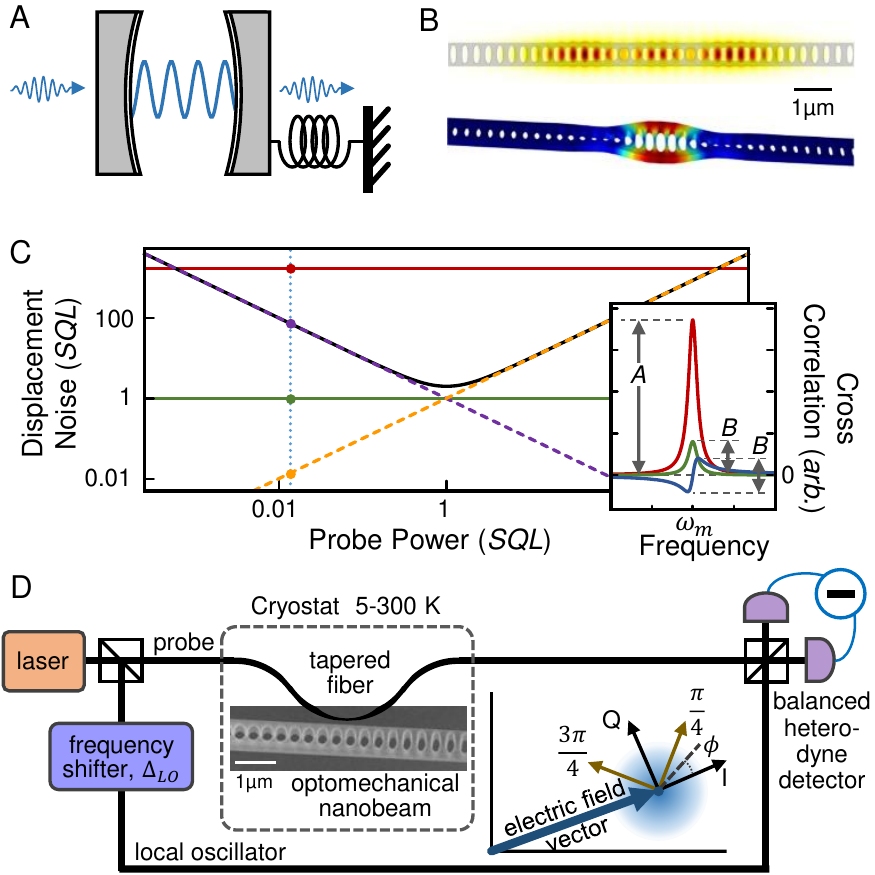}
	\end{center}
	\caption{Measurement backaction in an optomechanical system. (A) Effective optomechanical system for device pictured in panel B, a harmonically bound mirror forming one end of an optical cavity. (B)  Simulated electric field distribution (top), simulated displacement field distribution (bottom) of a suspended Si$_3$N$_4$ nanobeam optomechanical crystal. (C) Quantum limited displacement sensing involves the phase-quadrature shot-noise measurement floor (dashed purple) and the quantum measurement backaction due to radiation pressure shot noise (dashed orange).  The product of these two fundamental noise sources (green) is constant as dictated by a Heisenberg measurement -- disturbance uncertainty relation.  The sum of the two fundamental noise sources (black) is minimized at the standard quantum limit (SQL).  Brownian motion of the mechanical resonator is shown in red.  Our experiment takes place in the weak probe power regime (dotted blue) where measurement backaction is small.  Inset: Optomechanical cross correlations. The thermal correlation, $\mathrm{Re}\{S_{\frac{\pi}{4},\frac{3\pi}{4}}(\omega)\}$, is shown in red. The quantum correlation, $S_{0,\frac{\pi}{2}}(\omega)$, is shown in blue (real part) and green (imaginary part).  The ratio of the thermal motion peak, $A$, to the quantum correlation size, $B$, allows us to extract the temperature of the device.  (D) Experimental setup. Inset: Coordinate system for optical quadratures.}
\end{figure}

	Our system consists of a nanophotonic cavity coupled to a nanomechanical resonator, where vibrations of the mechanical mode modulate the optical resonance frequency and radiation pressure from an optical probe perturbs the mechanics (Fig.~1A,B)).  Considering the dynamics and noise in this optomechanical system reveals a simple picture of the origin of the measurement backaction and quantum correlations.  The nanomechanical resonator is driven to vibrate randomly in a band of frequencies around its resonance in response to thermal forces from its environment. This Brownian motion modulates the optical resonance frequency, imprinting phase modulation on a resonant optical probe.  Along with this displacement measurement, the random quantum intensity fluctuations of the probe drive the mechanics with so-called radiation pressure shot noise (RPSN), constituting the quantum measurement backaction.  The motion from backaction is also imprinted as phase fluctuations on the output light.  If $\delta X_I$ and $\delta X_Q$ are the quantum fluctuations of the optical amplitude and phase quadratures of the input probe, then, for a probe resonant with the optical cavity, the optomechanical interaction leaves $\delta X_I$ unchanged while $\delta X_Q \rightarrow \delta X_Q + \alpha \delta F_{\mathrm{th}} +\beta \delta X_I$, where the first term represents shot noise, the second represents the transduction of thermal force fluctuations, $\delta F_{\mathrm{th}}$, and the third represents the effect of RPSN.  Thus, a quantum correlation is established between amplitude and phase fluctuations when $\delta X_Q$ acquires a term proportional to $\delta X_I$.  Such correlations are the basis of phenomena including the generation of optomechanically squeezed light~\cite{Brooks2012,SafaviNaeini2013,Purdy2013b}, optomechanical Raman sideband asymmetry~\cite{SafaviNaeini2012,Lecocq2014,Weinstein2014,Purdy2015,Underwood2015,Sidhar2016,StamperKurn2014}, and the application of correlated squeezed light to enhance displacement sensing below the standard quantum limit~\cite{Unruh1983}.  Several methods have been explored to directly measure quantum-backaction-induced correlations~\cite{Heidmann1997,Verlot2009,Borkje2010,Purdy2013} (our method is similar to the proposal of Ref.~\cite{Borkje2010}), but the simplicity and systematic error rejection of the measurements described below allow us to resolve correlations that, for a room temperature device, are many thousands of times smaller than uncorrelated noise sources.

	After passing through the optomechanical system, the probe laser acquires sidebands (Fig.~2A) from the optomechanical interaction.  These sidebands can be thought of as the Stokes and anti-Stokes Raman scattering generated when a probe photon is down- or up- shifted in energy while a vibrational quanta from the mechanical resonator is created or annihilated.  In this picture quantum correlations manifest as an asymmetry between the sidebands~\cite{Khalili2012,Weinstein2014}.  However, it is most natural to consider our heterodyne detected optical signal in terms of fluctuations of optical quadratures, $\delta X_{\phi}=\cos(\phi) \delta X_I + \sin(\phi) \delta X_Q$, for quadrature angle $\phi$.  While quantum signals are not immediately evident in the spectrum of any individual quadrature, we can compute correlation spectra  of the form $S_{\phi_1,\phi_2}(\omega)=\langle\delta X_{\phi_1}^{*}(\omega) \delta X_{\phi_2}(\omega)\rangle$, normalized to shot noise (see supplementary materials~\cite{Supp}).  We designate the amplitude -- phase cross correlation, $S_{0,\frac{\pi}{2}}(\omega)$, as the quantum correlation (Fig.~2G). 
\begin{eqnarray} 
	S_{0,\frac{\pi}{2}}(\omega)=\beta(\omega)= D \chi_{\mathrm{m}}(\omega)
\end{eqnarray}
$D=2 \hbar \epsilon \bar{N} G^2 \kappa/(\kappa^2/4+\omega^2)$ represents the strength of the optomechanical transduction, with $\epsilon$ representing the effects of optical loss and finite detection efficiency, $\bar{N}$ the average intracavity photon occupation, $G=\frac{d\omega_c}{d x}$ the optomechanical coupling constant, and $\kappa$ and $\omega_c$ the optical decay rate and resonance frequency.  The quantum correlation represents a measurement of the linear response of the mechanics to the applied force of RPSN, characterized by a mechanical susceptibility, $\chi_{\mathrm{m}}(\omega)=1/(m(\omega_{\mathrm{m}}^2-\omega^2-i \Gamma_{\mathrm{m}} \omega))$, where $\omega_{\mathrm{m}}$, $m$, and $\Gamma_{\mathrm{m}}$ are the mechanical resonance frequency, effective mass, and damping rate.  Hence, the quantum correlation is complex with real and imaginary part following the dispersive and dissipative parts of a damped harmonic oscillator response, respectively.  Thermal signals and measurement noise backgrounds, all of which are uncorrelated between amplitude and phase, average to zero.  The characteristic scale of the quantum correlation is set by the Heisenberg measurement-disturbance uncertainty relation and is equivalent to the scale of mechanical vacuum fluctuations, when expressed as an effective displacement noise (as opposed to the shot-noise-normalized form of Eq.~1). In these units the quantum correlation is independent of probe power, evident by noting that $S_{0,\frac{\pi}{2}}$ is fundamentally a product of backaction-driven motion and shot noise, similar to the Heisenberg limited measurement backaction -- shot noise imprecision product shown in Fig.~1C.  For the room temperature data of Fig.~3A,B, the RPSN-driven motion is estimated to be about $3\times 10^{-6}$ smaller than the thermal motion, while the quantum correlation is about $2 \times 10^{-4}$ smaller than the thermal motion signal.

\begin{figure}
	\begin{center}
 \includegraphics[width=3.1 in]{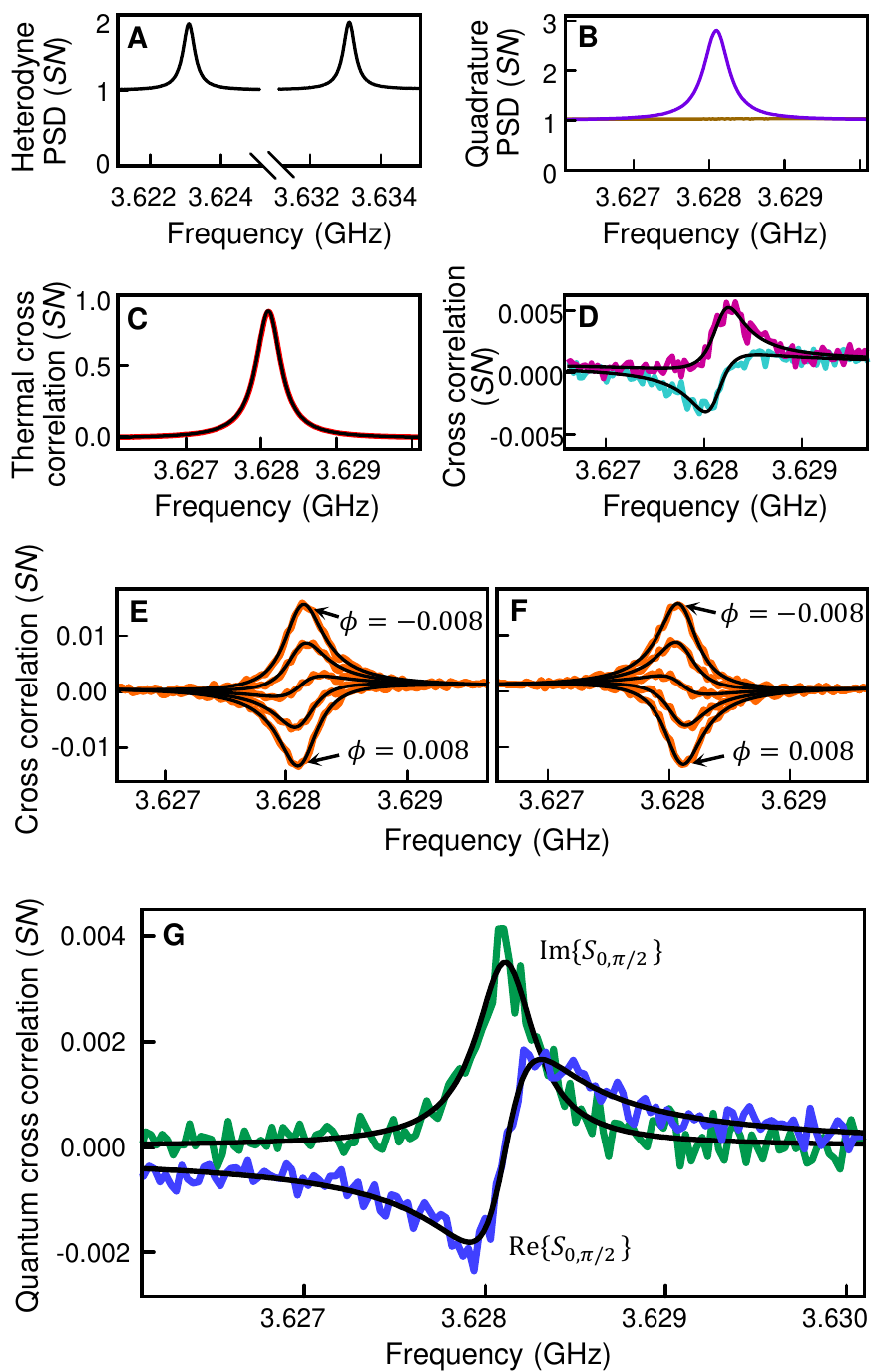}
	\end{center}
 \caption{Optical correlation spectra.  (A) Power spectral density (PSD) of heterodyne photodetector signal, normalized to shot noise (SN).  Stokes and anti-Stokes mechanical sidebands are shown.  (B) PSD of the amplitude quadrature ($S_{0,0}(\omega)$, brown) displaying shot noise and the phase quadrature ($S_{\frac{\pi}{2},\frac{\pi}{2}}(\omega)$, purple) showing the mechanical signal and shot noise.  (C) Spectrum of thermal cross correlation, $\mathrm{Re}\{S_{\frac{\pi}{4},\frac{3\pi}{4}}(\omega)\}$, (red), Lorentzian fit (black).   (D) $\mathrm{Re}\{S_{0,\frac{\pi}{2}}(\omega)\}$ for $\Delta_p$ differing by about 0.005 $\kappa$ (cyan, magenta), fits (black). (E),(F) Cross correlation spectra, $\mathrm{Re}\{S_{\phi,\phi+\frac{\pi}{2}}(\omega)\}$, for $\Delta_{\mathrm{LO}}/2\pi=+5$ MHz (E), and  $\Delta_{\mathrm{LO}}/2\pi=-5$ MHz (F).  From top to bottom curve of each panel, $\phi$ varies from -0.008 to 0.008 in 0.002 steps (orange), fits (black). (G) Spectrum of quantum cross correlation, real part ($\mathrm{Re}\{S_{0,\frac{\pi}{2}}(\omega)\}$, blue), imaginary part ($\mathrm{Im}\{S_{0,\frac{\pi}{2}}(\omega)\}$, green), fits (black).  All traces are generated from the same 300~s of time domain data. $T=22$~K.}
\end{figure}

	To measure thermal signals, we could simply compute the power spectrum of the phase quadrature and subtract off any unwanted noise backgrounds.  However, a more elegant approach to isolate the thermal signal is to consider the cross correlation:
\begin{eqnarray} 
	S_{\frac{\pi}{4},\frac{3\pi}{4}}(\omega)=2 D \mathrm{Im}\{\chi_m(\omega) \} (n_{\mathrm{th}}(\omega)+1/2)+ i \mathrm{Im}\{S_{0,\frac{\pi}{2}}(\omega)\} 
\end{eqnarray}
where $n_{\mathrm{th}}(\omega)=1/(\exp(\hbar \omega/k_{\mathrm{B}} T)-1)\approx k_\mathrm{B} T / \hbar \omega_{\mathrm{m}}$ is the average mechanical occupation of the mechanical bath at temperature $T$, coupled the resonator.  We designate $\mathrm{Re}\{S_{\frac{\pi}{4},\frac{3\pi}{4}}(\omega)\}$ as the thermal correlation. It retains the thermal motion contribution of the phase quadrature power spectrum, but has zero background (Fig.~2B,C).  The form of the thermal correlation is governed by a fluctuation-dissipation theorem~\cite{Callen1951}, evidenced in Eq.~2 where the imaginary part of the mechanical susceptibility appears linking the mechanical dissipation to thermal fluctuations.  $\mathrm{Im}\{S_{\phi,\phi+\frac{\pi}{2}}(\omega)\}$ is independent of $\phi$ and is purely quantum when the probe -- cavity detuning, $\Delta_p$, is zero.  From Eq.~2 we have:
\begin{eqnarray}
  \frac{\mathrm{Re}\{S_{\frac{\pi}{4},\frac{3\pi}{4}}(\omega)\}}{\mathrm{Im}\{S_{\frac{\pi}{4},\frac{3\pi}{4}}(\omega)\}}=\coth \left(\frac{\hbar \omega}{2 k_{\mathrm{B}} T} \right)
\end{eqnarray} 
This relationship allows us to measure temperature from the ratio of components of correlation spectra without detailed knowledge of the optomechanical device parameters, detection efficiency, and mechanical and optical susceptibilities.  Our Brownian motion thermometer is similar in concept to other types of noise measurement and thermometry, such as Johnson noise in electrical systems, blackbody radiation, or optical fiber Raman scattering, that can be calibrated with fundamental quantum noise~\cite{Dakin1985,Pekola1994,Migdall1999,Spietz2003}.

	Our measurements employ a Si$_3$N$_4$ optomechanical crystal~\cite{Davanco2014,Grutter2015} (Fig.~1B), a suspended Si$_3$N$_4$ nanobeam patterned with a series of holes that produces a mechanical resonance, frequency $\omega_{\mathrm{m}}/2\pi= 3.62$~GHz, which varies over 10 MHz from cryogenic to room temperature, and an optical resonances near 991~nm, coupled at a rate $g_0=G \sqrt{\hbar/2 m \omega_{\mathrm{m}}}\approx 2 \pi \times 70$~kHz. The mechanical decay rate varies from $\Gamma_{\mathrm{m}}/2\pi=1.4$~MHz at room temperature to an order of magnitude lower at cryogenic temperatures.   We choose the optomechanical crystal geometry~\cite{Eichenfield2009} for its large optomechanical coupling at high mechanical frequency, facilitating high signal to noise readout of mechanical motion at relatively low $n_{\mathrm{th}}(\omega_m)$ (tens to thousands from cryogenic to room temperature), as well as for its potential for device integration and atmospheric pressure operation.  Probe light is evanescently coupled to the optical resonance through a tapered optical fiber, and detected with $\epsilon \approx 0.1$.  While the intrinsic optical decay rate is $2 \pi \times 2.3$~GHz, we intentionally overcouple the device to the tapered fiber to achieve an overall optical decay rate $\kappa/2\pi\approx10$~GHz.  All of our measurements are in the weak probe regime with optomechanical cooperativity $C=4 \bar{N} g_0^2/\kappa\Gamma_{\mathrm{m}}$ typically in the range of $0.01$ to $0.1$, minimizing the effects of dynamical backaction and RPSN induced motion~\cite{Purdy2013}. Classical noise on the probe laser interacting with the mechanics could induce additional optomechanical cross correlations.  However, excess laser noise is measured to be far below the shot noise level, precluding this possibility.  The probe wavelength is actively stabilized near optical resonance, with a residual detuning $\Delta_{\mathrm{p}}/\kappa$ at the percent level.  (See supplementary materials for more experimental details~\cite{Supp}.)

	To obtain a purely quantum correlation signal at large $n_{\mathrm{th}}$, we must carefully understand the effects of finite probe detuning and imperfect quadrature determination.  Residual probe detuning mixes the amplitude and phase quadratures at the output of the cavity (Fig.~2D), adding to $S_{0,\frac{\pi}{2}}(\omega)$, a thermal contribution that becomes dominant when $\Delta_p/\kappa > 1/n_{\mathrm{th}}$.  We counter the effects of finite detuning by rotating the analyzed pair of orthogonal quadratures by $\phi\sim\Delta_p/\kappa$.  This postprocessing undoes the cavity-induced optical quadrature rotation for the real component of the cross correlation (Fig.~2E).  By toggling the sign of the heterodyne -- probe detuning, $\Delta_{\mathrm{LO}}\rightarrow -\Delta_{\mathrm{LO}}$, we perform a type of lock-in detection, where the quantum correlation inverts (Fig.~2E,F), while thermal correlations resulting from imperfections and dispersion in the optical quadrature determination remain unchanged.  With these two tools we are able to reject thermal signals from the real component of the quantum correlation at all temperatures and from the imaginary component of the cross correlation only at low temperatures, as the demands on the stability of $\Delta_p$ grow with increasing $n_{\mathrm{th}}$ (see supplementary materials for details~\cite{Supp}).

\begin{figure}
\begin{center}
\includegraphics[width=3.1 in]{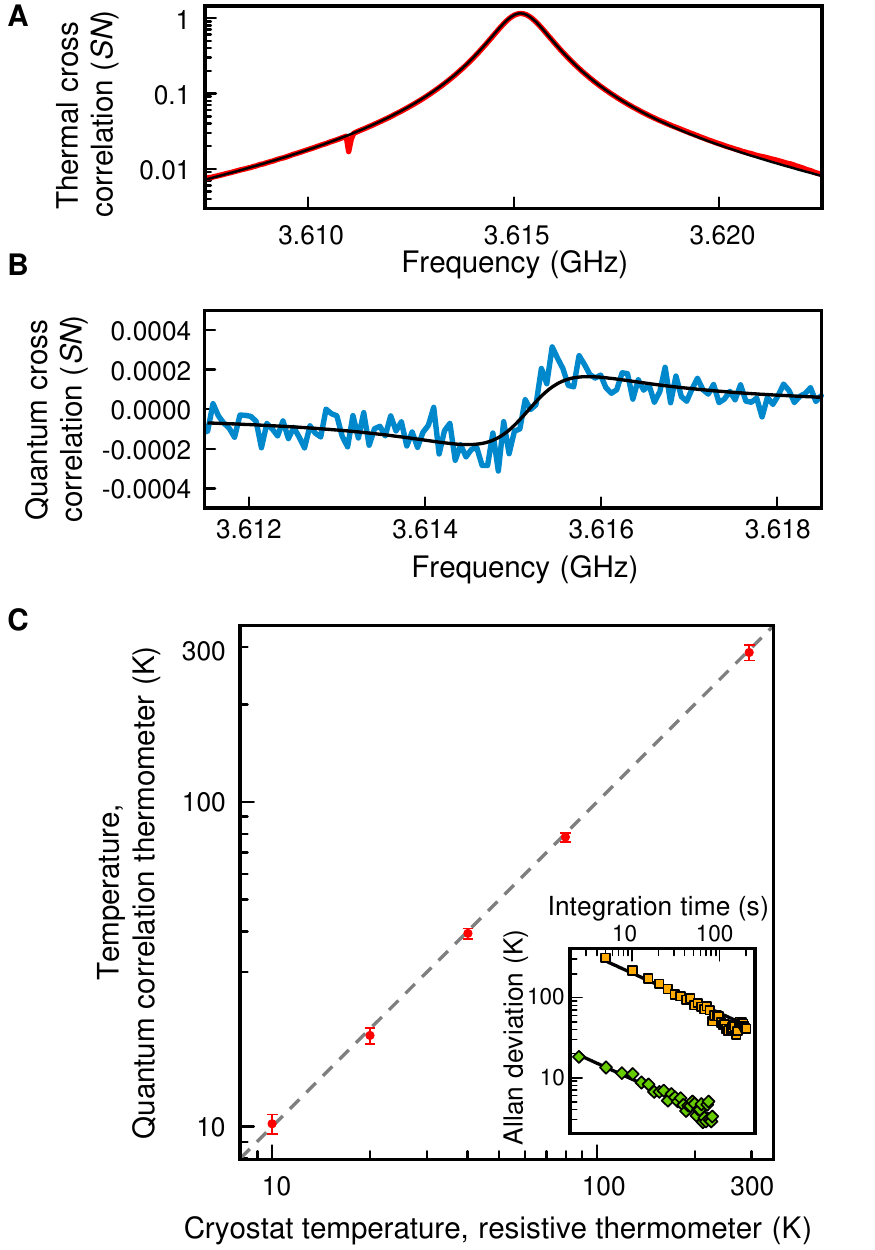}
\end{center}
\caption{Quantum correlation thermometry.  (A), (B) Quadrature cross correlations are shown for operation at room temperature, $T=294$~K, and atmospheric pressure, $\mathrm{Re}\{S_{\pi/4,3\pi/4}(\omega)\}$, (panel A, red), $\mathrm{Re}\{S_{0,\pi/2}(\omega)\}$, (panel B, blue).  Solid black curves are fits to data. In panel A the sharp feature at 3.611~GHz is an artifact of the heterodyne local oscillator, and the very small, broad deviation from the Lorentzian fit around 3.622~GHz is most likely another weakly coupled mechanical mode.  The spectral band used for temperature determination does not include either of these regions.  (C) Temperature measured via quantum correlation thermometry is plotted (red circles) as the system temperature is varied.  Dashed gray line represents agreement between quantum correlation thermometry and a resistive thermometer measuring the sample temperature. The statistical error bars represent the estimated standard deviation. The inset shows the Allan deviation of the temperature measurement for 294~K (orange squares) and 40~K (green diamonds).  Fit curves (black) correspond to 640~K /$\sqrt{\mathrm{Hz}}$ and 30~K /$\sqrt{\mathrm{Hz}}$ for the 294~K and 40~K data respectively.}
\end{figure}

	Fig.~3B shows the quantum correlation obtained for our device at room temperature.  Given our ability to resolve $\mathrm{Re}\{S_{0,\frac{\pi}{2}}(\omega)\}$ over a wide temperature range, we look to apply it to Brownian motion thermometry.  We could formally apply a Kramers-Kronig transformation to express $\mathrm{Im}\{S_{\frac{\pi}{4},\frac{3\pi}{4}}(\omega)\}$ in Eq.~3 terms of $\mathrm{Re}\{S_{0,\frac{\pi}{2}}(\omega)\}$, reconstructing the imaginary part of the mechanical susceptibility from the real part, assuming $\chi_{\mathrm{m}}(\omega)$ is analytic.  We use a simple approximation of this idea, valid for a weakly damped simple harmonic oscillator.   We fit curves such as those shown in Fig.~3A,B to obtain the height of the Lorentzian thermal correlation, $A$, and the peak-to-peak value of the real part of the quantum correlation, $B$, which is also equal to the peak height of the imaginary quantum correlation (Fig.~1C).  The ratio $A/2 B\approx k_{\mathrm{B}} T/\hbar \omega_{\mathrm{m}}$, is our approximate measure of the absolute temperature of the mechanical resonator.
		
		Fig.~3C shows agreement at the few percent level between this quantum correlation thermometry and the independently measured cryostat temperature over the range of 10~K to 294~K. Allan deviations for the measurement are shown in the inset of Fig.~3C. At higher temperatures, the statistical noise in determining the quantum correlation is dominated by thermal fluctuations, while at lower temperatures the shot-noise measurement floor is dominant.  Each data point in Fig.~3C includes between several hundred and one thousand seconds of integration time.  The dominant systematic effect at low temperatures is self-heating due to optical absorption of the probe light, spurred by  low thermal conductivity of silicon nitride thin films, which drops dramatically below a few tens of Kelvin~\cite{Zink2005}.  To account for this effect, we probe our device with a range of optical power and extrapolate the temperature at zero probe power.  We have applied this correction to the data points below 40~K in Fig.~3C.  At higher temperatures the self-heating is negligible compared to the statistical uncertainty.
	
	We have demonstrated a proof-of-principle Brownian motion thermometer calibrated with optomechanical quantum correlations. We use well understood optical quantum force fluctuations as a standard reference scale for measuring thermal force fluctuations.  With improved devices, for example the increased optomechanical coupling and thermal conductivity afforded by semiconductor materials~\cite{Eichenfield2009} and increased mechanical resonance frequency~\cite{Han2015}, we believe this technique could readily attain metrologically useful accuracy levels over a wide temperature range, and become relevant as an on-chip, photonic, independent primary thermometry method~\cite{Flowers2004}.  More generally, this first glimpse of quantum effects in a mechanical system strongly coupled to a room temperature environment represents a step toward room temperature quantum sensing and information applications. 

\textbf{Acknowledgments:}  We thank Krishna Coimbatore Balram and Marcelo Davan\c{c}o for technical assistance and acknowledge useful discussions with Zeeshan Ahmed, Nikolai Klimov, and Gregory Strouse.\\

\nocite{Jayich2012,Giovannetti2001} 

\bibliography{OMQC}

\clearpage

\onecolumngrid

\begin{center}

\large{\textbf{Supplementary Materials for:}}\\
\large{\textbf{Observation of Optomechanical Quantum Correlations at Room Temperature}}\\
\bigskip

\normalsize{T. P. Purdy,$^{1\ast}$ K. E. Grutter,$^{2}$ K. Srinivasan$^{2}$, J. M. Taylor$^{1,3}$}\\
\medskip

\textit{
\small{$^{1}$Joint Quantum Institute, National Institute of Standards and Technology,Gaithersburg, MD 20899 USA}\\
\small{$^{2}$Center for Nanoscale Science and Technology, National Institute of Standards and Technology,Gaithersburg, MD 20899 USA}\\
\small{$^{3}$Joint Center for Quantum Information and Computer Science, University of Maryland,College Park, Maryland 20742, USA}\\
\normalsize{$^\ast$thomas.purdy@nist.gov.}\\
}
\medskip

\end{center}

\twocolumngrid

\beginsupplement

\section{Experimental Methods}

\subsection{Optical Setup}
	A diagram of the experimental setup is given in Fig.~S1.  We employ a low noise Ti:sapphire laser to generate the probe, frequency $\omega_p$, and heterodyne local oscillator (LO) beams.  The LO is split off from the probe and deflected by two successive acousto-optic modulators (AOM) causing a net shift in the LO frequency of either $\Delta_{\mathrm{LO}}/2\pi=+5$~MHz or -5~MHz depending on the selected diffraction order.  The probe is frequency modulated at 2.6~GHz with an electro-optic modulator (EOM) used to generate an error signal for active stabilization of the laser to optical resonance.  For some experiments another weak modulation is applied at a frequency near the mechanical resonance to serve as a calibration tone.  The probe light is evanescently coupled into the optomechanical cavity with a dimpled, tapered optical fiber.  After interacting with the cavity, the probe is recombined with the LO on a 50:50 fiber beam splitter and detected with a balanced photodetector.

\begin{figure*}
\includegraphics[]{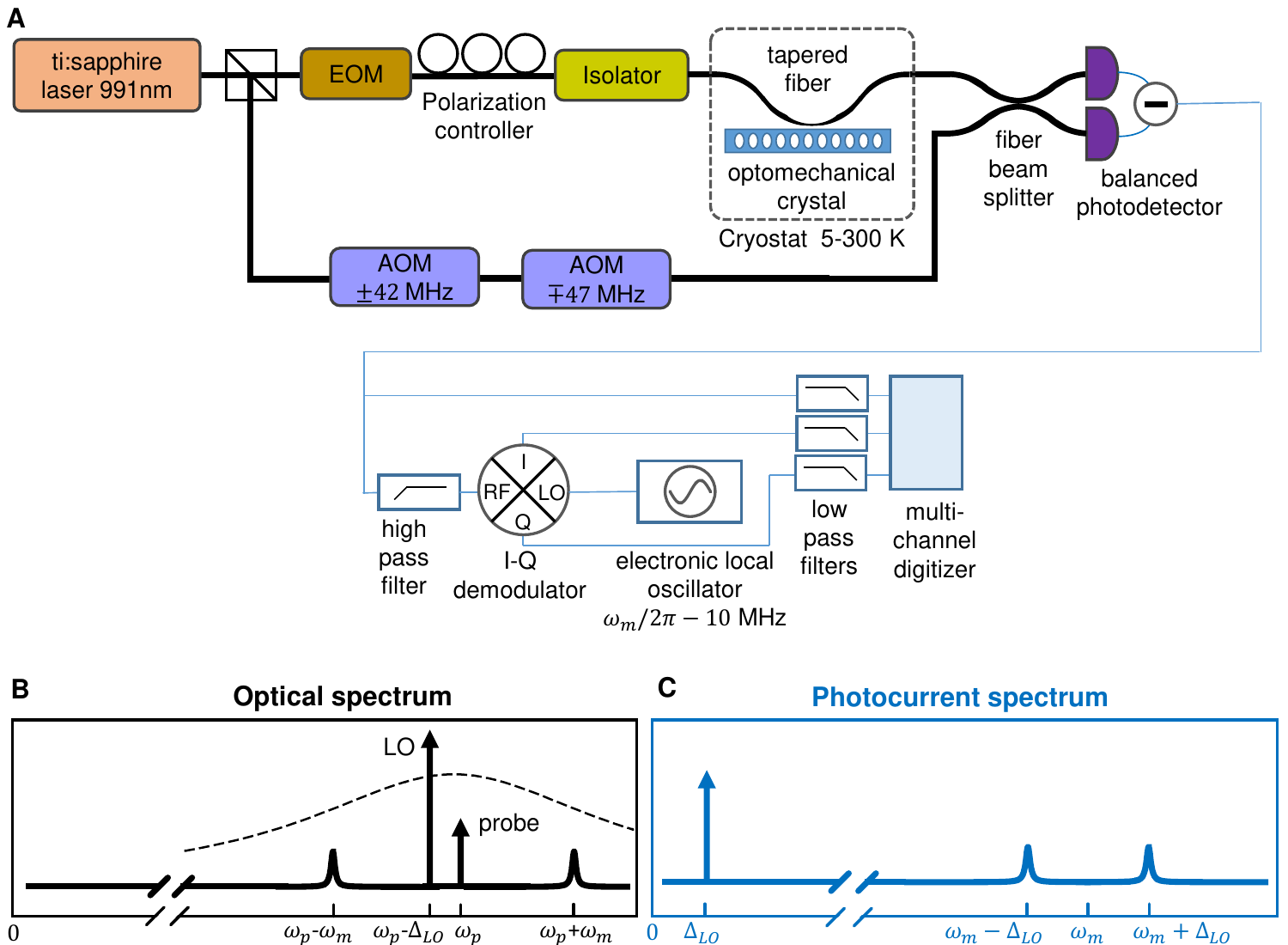}

\caption{Experimental Setup. (A) Laser and detection electronics diagram.  (B) Spectral content of optical signal at photodetector (solid black).  Optical cavity Lorentzian response (dashed black) (C) Spectral content of heterodyne photocurrent signal (blue). }
\end{figure*}
	
\subsection{Electronic and Digital Data Processing}
		The signal from the photodetector is split into a low frequency component containing the beat signal at $\Delta_{\mathrm{LO}}$ between the LO and probe carrier, and a high frequency component with the optomechanical sidebands at $\omega_{\mathrm{m}} \pm \Delta_{\mathrm{LO}}$ (Fig.~S1C).  The high frequency components are mixed down to frequencies within the range of our digitizer (30~MHz capture bandwidth).  The carrier, the in-phase, and the out-of-phase signals from the mixer are all captured with a multichannel digitizer.  From the carrier channel, the phase of the beat note as a function of time is computed.  The cosine and sine of this phase is multiplied by the complex time domain optomechanical sideband signal to digitally demodulate the signal and generate two orthogonal optical quadratures.  By capturing the carrier beat note and accounting for its phase drift, we do not require long term interferometric stability between the probe and LO beam paths.
		
		We apply a frequency dependent correction to account for the gain of the photodetector and ensuing electronics.  The magnitude of the correction is obtained by measuring the transduction of optical shot noise through the electronics, which should have a flat spectrum.  The phase response is measured in a separate experiment where we phase modulate the probe light with a comb of modulation frequencies in a band around the mechanical resonance and compare the detected relative quadrature phase of each comb tooth.  It is particularly important to correct phase errors because the effect of residual higher-order electronic phase dispersion is to rotate the apparent optical quadrature of signals in a frequency dependent fashion.  This electronic dispersion can cause spurious features in the amplitude -- phase cross correlation. If a linear-in-frequency apparent quadrature rotation is applied to Lorentzian distributed thermal phase noise, the result is a spurious dispersive feature that might easily be confused for a quantum correlation.
	
		Our postprocessing data analysis allows us to accurately identify the pair of orthogonal quadratures that contain purely quantum correlations. Any small detuning of the probe from cavity resonance rotates the quadratures of the output light by an amount on the order of $2 \Delta_{p}/\kappa$ (Fig.~2D of main text).  While we actively stabilize the probe near optical resonance, a small detuning remains with $2 \Delta_{\mathrm{p}}/\kappa$ at the percent level.  We map out the quadrature cross correlation spectra, $\mathrm{Re}\{S_{\phi,\phi+\frac{\pi}{2}}(\omega)\}$, as shown in Fig.~2E,F of the main text. The quadrature pair of maximum quantum correlation signal is rotated, but to first order in the $\Delta_{p}/\kappa$, its magnitude is not affected.  However, in this region the thermal component of the cross correlation varies linearly with $\phi$, and is exactly nulled when the quantum correlation is maximized.  The thermal and quantum correlations become similar in magnitude for deviations of $\phi$ on the order of $1/n_{\mathrm{th}}$ from the optimum angle. Rather than attempt to accurately measure the cavity -- probe detuning, we fit a series of curves of $\mathrm{Re}\{S_{\phi,\phi+\frac{\pi}{2}}(\omega)\}$ such as those shown in the main text Fig.~2E,F for several values of $\phi$ around $\phi=0$ to find the quadratures with no thermal contribution.   This procedure effectively separates the small quantum correlation from the much larger thermal signal.  For all the data presented in the main text, we redefine $\phi=0$ to correspond to this shifted minimum of thermal signal, and use this shifted coordinate for subsequent determination of quantum and thermal correlations.  Because $\mathrm{Im}\{S_{\phi,\phi+\frac{\pi}{2}}(\omega)\}$ is independent of $\phi$, this procedure cannot be used to separate the imaginary component of the quantum correlation from the imaginary thermal contribution in the case of $\Delta_p\neq 0$.   At higher temperatures the requirement on laser lock stability becomes more stringent, and so we are only able to extract the purely quantum component of the imaginary cross correlation at cryogenic temperatures. 
		
			We can further separate thermal and quantum correlations in both the real and imaginary components of the cross correlation by making measurements with opposing values of LO detuning, $\Delta_{\mathrm{LO}}/2\pi=5$~MHz or $\Delta_{\mathrm{LO}}/2\pi=-5$~MHz.  By toggling the sign of the heterodyne detuning, $\Delta_{\mathrm{LO}} \rightarrow -\Delta_{\mathrm{LO}}$, we invert the phase modulation signal, $\delta X_{Q} \rightarrow -\delta X_{Q}$ while leaving the amplitude modulation as well as any residual detector electronics dispersion unchanged.  Thus the quantum amplitude -- phase correlation inverts, while spurious electronic-dispersion-induced thermal noise correlations remain unchanged.  The results of toggling the LO detuning are shown in Fig.~2E,F of the main text.  The dispersive quantum correlation inverts between positive and negative LO detuning data and their magnitudes agree to within the statistical uncertainty of the fits.  At higher temperatures, where the ratio of thermal to quantum correlations is larger, significant differences in the amplitude -- phase cross correlation signal are present between positive and negative LO detunings if the phase calibration described above is not performed accurately.  To combat any residual phase calibration error, all of the spectra used for thermometry are averaged over data taken with opposing LO detunings.  Quantum correlations are calculated as half the difference of the two LO detunings, and thermal correlations are calculated as half the sum of the two LO detunings.

\subsection{Mechanical linewidth}
   The mechanical linewidth is a strong function of temperature.  Fig.~S2 shows the measured linewidth to be a monotonic function of temperature.  The room temperature data displayed in the main text was taken at atmospheric pressure, where the mechanical linewidth is slightly larger.  All other data were taken under high vacuum conditions.   

\begin{figure}
\includegraphics[]{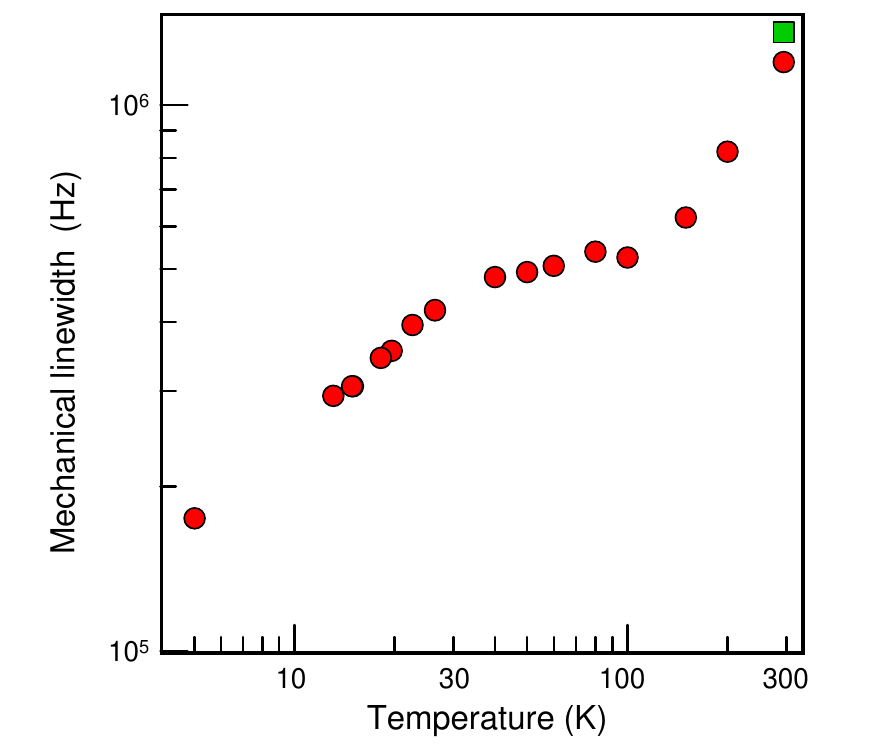}

\caption{Mechanical linewidth.  Measured linewidths under high vacuum are plotted as red circles.  Green square data point was taken at room temperature and at atmospheric pressure. Typical reproducibility with temperature cycling is at the percent level.}
\end{figure}

\subsection{Laser noise}

Classical laser noise can drive mechanical motion and create classical amplitude -- phase optomechanical cross correlations~\cite{Jayich2012,Weinstein2014,Sidhar2016}.  Our ti:sapphire laser has very low classical noise at frequencies around the mechanical resonance frequency away from the carrier.  We assess the laser noise before it interacts with the optomechanical cavity.  Excess amplitude noise is measured by directly detecting laser light on a high speed photodetector and looking at its noise in a frequency band near the mechanical resonance frequency.  Fig.~S3A shows a linear scaling of laser intensity noise power with laser power indicative of shot noise.  The linear scaling persists up to 1~mW of measured power.  Typical probe laser powers are only on the order of 10~$\mu$W, so we conclude that the amplitude noise of the laser is shot noise limited.   We next detect the probe laser with optical heterodyne, after passing the probe through an approximately 10~m long delay line to decorrelate noise common to the probe and heterodyne local oscillator.  The noise power as a function of optical quadrature from this test is shown in Fig.~S3B.  Deviations from shot noise are at the $10^{-3}$ level, and do not vary substantially within a few tens of MHz around the mechanical resonance frequency.  This test indicates that all optical quadratures are shot noise limited.  We also display the cross correlation, $\mathrm{Re}\{S_{\phi,\phi+\frac{\pi}{2}}(\omega)\}$, obtained during this test and observe deviations from zero only at the level of $10^{-3}$ of shot noise.  Any classical noise on the laser will not contribute significantly to the data presented in the main text.

\begin{figure}
\includegraphics[]{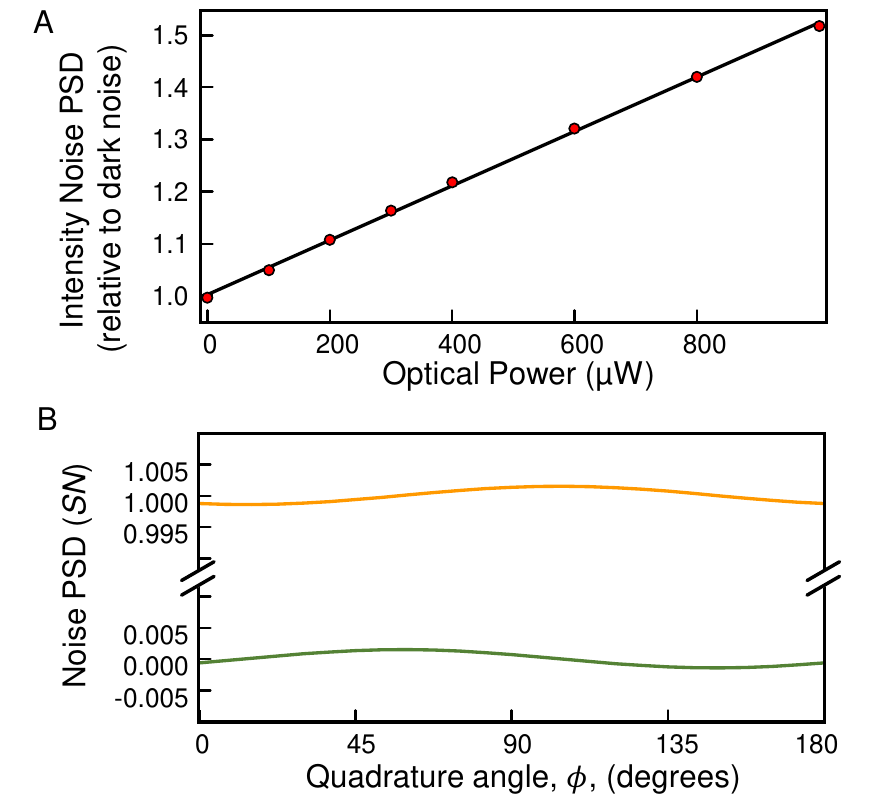}

\caption{Laser noise.  (A) Measured intensity noise power spectral density (PSD) (red circles), linear fit (black).  (B) Measured noise in optical quadratures relative to shot noise (SN), orange.  Measured noise in optical quadrature cross correlations, $S_{\phi,\phi+\frac{\pi}{2}}(\omega)$, green.  All data are for light that has not interacted with the optomechanical cavity and are averaged over a few MHz band near the mechanical resonance frequency.}
\end{figure}

\subsection{Photodetector linearity}

	With a typical LO power of 2~mW, we do see small effects of photodetector nonlinearity.  Intermodulations at $\omega_{\mathrm{m}}\pm 2\Delta_{\mathrm{LO}}$ are present at the -30~dB to -40~dB level relative to the expected heterodyne sidebands at  $\omega_{\mathrm{m}}\pm \Delta_{\mathrm{LO}}$.  A slight saturation in the broadband response of the photodetector at the percent level is also observed. The effect of this nonlinearity is to add a nonzero floor to the cross correlation spectrum at the level of $10^{-4}$ to $10^{-3}$ of shot noise.  Because we are able to obtain statistically consistent results with optomechanical quantum correlation thermometry while varying the LO power by more than a factor of two, we do not believe that photodetector nonlinearity is significantly impacting our results.  However, we believe that such nonlinearity may become an important systematic concern if higher accuracy thermometry is pursued in the future.

\subsection{Detection efficiency}
	There are several sources of loss between intracavity photons and our detector.  We estimate these losses and provide typical values. Only a fraction of intracavity photons, $\epsilon_1$, are emitted back into the fiber taper propagating toward the detector due to intrinsic cavity loss and the two-sided nature of the cavity. $\epsilon_1=\kappa_{\mathrm{out}}/\kappa\approx 0.38$, where $\kappa_{\mathrm{out}}$ is the decay of the cavity into output port of the fiber taper.  Photons are lost in the fiber taper, fiber-to-fiber coupling, and beam splitter, yielding $\epsilon_2 \approx 0.5$.  The photodetector conversion efficiency is $\epsilon_3 \approx 0.80$.  We account for the dark noise floor of the photodetector as an effective quantum efficiency of $\epsilon_4 \approx 0.6$.  These effects yield an overall detection efficiency $\epsilon=\epsilon_1 \times \epsilon_2 \times \epsilon_3 \times \epsilon_4 \approx 0.09$.  We include an additional factor of 0.5 in the effective detection efficiency to account for loss in signal to noise of heterodyne detection of any particular optical quadrature, compared to an ideal single quadrature (homodyne) detection scheme.    

\subsection{Cryostat temperature}
	Our experiments take place in a low-vibration, closed-cycle cryostat with base temperature of 4.5~K.  The silicon die supporting the Si$_3$N$_4$ optomechanical crystal is attached with low-temperature vacuum grease to a variable temperature stage inside the cryostat.  The stage temperature is monitored with a Cernox resistance temperature detector, with typical long-term absolute stability of better than 0.18~K according the manufacturer (Certain commercial products are identified in this paper to foster understanding. Such identification does not imply recommendation or endorsement by the National Institute of Standards and Technology, nor does it imply that the materials or equipment identified are necessarily the best available for the purpose.).  A resistive heater is used to actively stabilize the stage temperature with a short term stability of a few mK.  For measurements below room temperature, the cryostat is maintained under high vacuum, and the device is otherwise surrounded by a radiation shield at a temperature of about 30~K.  For the room temperature measurements the cryostat is held at atmospheric pressure, and the radiation shield is also at room temperature.

\subsection{Optical absorption self-heating}
	We use several methods to assess and correct for self-heating due to absorption of laser light.  While holding the cryostat at a constant temperature, we vary the optical power used to probe the sample.  We use either the relative fractional area of the Lorentzian thermal correlation or the mechanical linewidth variation with temperature as a secondary, relative thermometer of the mechanical temperature.  Both methods give consistent results.  We have also attempted full quantum correlation thermometry measurements at several optical probe powers.  Such measurements also give consistent results, but the statistical uncertainty at low power is much larger than the other techniques.

\subsection{Laser wavelength stabilization}
	We actively stabilize the laser wavelength to the optical resonance.  We generate an error signal by weakly phase modulating the probe laser at 2.6~GHz.  This modulation is detected in our heterodyne detector along with the optomechanical signal.  The phase modulation frequency is chosen to be well separated in frequency from any optomechanical signals.  We split the heterodyne photodetector signal into a low frequency component containing the probe carrier -- LO beat note, and a high frequency component containing the phase modulation sideband -- LO beat notes.  The high frequency component is first demodulated with an electronic local oscillator at the phase modulation frequency, and then demodulated again at $\Delta_{\mathrm{LO}}$ by mixing with a phase shifted version of the low frequency photodetector signal.  If proper phases are chosen for each demodulation step, a dispersive error is generated.  We feed this signal back to a piezoelectric transducer controlling the laser wavelength and achieve a feedback bandwidth of about 1~kHz.  Errors in the demodulation phases, optical power drift, and drifts in analog electronics limit the long term stability of this locking procedure to $\Delta_{\mathrm{p}}/\kappa$ at the percent level.

\section{Theoretical Methods}
	We discuss the theory of a single optical cavity mode optomechanically coupled to a single mechanical resonator~\cite{Giovannetti2001,Borkje2010,Clerk2010}. We start from a Hamiltonian, $H_0$ describing the coherent dynamics of the system.
\[
H_0=\hbar \omega_c a^{\dag} a +\frac{p^2}{2 m} +\frac{m \omega_m^2 x^2}{2} + \hbar G x a^{\dag} a
\]
where $a$ is the optical cavity annihilation operator, $\omega_c$ is the optical resonance frequency, $x$ and $p$ are the mechanical resonator position and momentum operators, $m$ and $\omega_m$ are the mechanical resonator effective mass and oscillation frequency, and $G=d \omega_c/dx$ is the optomechanical coupling constant.  $g_0=G x_{zp}$ is the optomechanical coupling rate, where $x_{zp}=\sqrt{\hbar/2 m \omega_m}$ is the mechanical zero point motion.

\subsection{Heisenberg-Langevin equations of motion}
	From $H_0$ we can write a set of Heisenberg-Langevin equations of motion describing the coherent dynamics as well as the coupling of the mechanics to a thermal bath and the coupling of the optical cavity to its input and output modes.\\
\begin{align*}
	\dot{x}&=p/m\\
	\dot{p}&=-m \omega_m^2 x -\hbar G a^{\dag} a -\Gamma_m p +\delta F_{th}\\
	\dot{a}&=-i \omega_c a -i G x a -\frac{\kappa}{2} a +\sum_{i}\sqrt{\kappa_i} a_{in,i}\\
\end{align*}
$\delta F_{th}$ is the random Langevin force on the mechanics from its thermal bath, which drives Brownian motion.  $\Gamma_m$ is the mechanical damping rate to the thermal bath.  $\kappa$ is the optical decay rate.  $\kappa_i$ represents the optical decay rate into the $i$th port of the optical cavity, with $\sum_{i} \kappa_i=\kappa$. $a_{in,i}$ are the cavity input annihilation operators.    One port of the cavity is driven with a coherent state described by $a_{in,1}(t)=(\bar{a}_{in}+\xi_{in,1}(t))e^{-i \omega_p t}$, where $\bar{a}_{in}$ is the coherent state amplitude, $\omega_p$ is the frequency of the coherent laser drive.  All other ports are assumed to be driven only by optical vacuum fluctuations, $a_{in,i}=\xi_{in,i}$ for $i\neq1$. Here, $\xi_{in,i}$ is the input vacuum noise operator for the $i$th port.  We can then write $a=(\bar{a}+d(t))e^{-i \omega_p t}$, where $\bar{a}=\sqrt{\kappa_1}\bar{a}_{in}/(\kappa/2-i \Delta_p)$ is the intracavity coherent state amplitude and $\Delta_p=\omega_p-\omega_c$ is the detuning of the input laser away from optical resonance, including the effect of any constant shift in optical resonance frequency induced by radiation pressure. We assume $\bar{a}$ to be real.  $\bar{a}^2=\bar{N}$, where $\bar{N}$ is the average intracavity photon occupation number.  $d$ represents the small fluctuations in the intracavity field.  We linearize the equations of motion around this steady state, and solve them in Fourier space. \\
\begin{widetext}
\begin{gather}
	x(\omega)=\frac{\chi_m(\omega)}{1-i \hbar G^2 \bar{a}^2 \chi_m(\omega)(\chi_c(\omega)-\chi_c^*(-\omega))} \left(\delta F_{th}(\omega)-\hbar G \bar{a} \sum_{i} \sqrt{\kappa_i}\left(\chi_c(\omega)\xi_{in,i}(\omega)+\chi_c^*(-\omega)\xi_{in,i}^{\dag}(\omega)\right) \right)\\
	d(\omega)=\chi_c(\omega) \left(-i G \bar{a} x(\omega)+\sum_{i}\sqrt{\kappa_i} \xi_{in,i}(\omega)\right)
\end{gather}  
Here, $\chi_m(\omega)=1/(m(\omega_m^2-\omega^2-i\Gamma_m \omega))$ is the mechanical susceptibility and $\chi_c(\omega)=1/(\kappa/2-i(\omega-\Delta_p))$ is the optical cavity susceptibility.  To simply illustrate the essential physics of quantum correlations, we will present next only the results for the experimentally relevant case of $\Delta_p=0$.  In general, finite detuning modifies mechanical susceptibility by the optical spring and optical damping effects~\cite{Clerk2010}, both of which are small when $C=4 \bar{N} g_0^2/\kappa\Gamma_m \ll 1$ or $\Delta_p \ll \kappa$.  Both of these conditions are satisfied for our experiment.  Also, we have verified using the full solution of the equations of motion for $\Delta_p \neq 0$ that the dominant effect on the output light is to slightly rotate the output quadratures that contain the pure quantum correlation signal (Fig.~S4).  We undo this rotation in postprocessing our data as described above.  For simplicity we will also present only the solution for an intrinsically lossless optical cavity with one input port and one output port, employing the simplifying notation $a_{in,1}\rightarrow a_{in}$, $\xi_{in,1}\rightarrow \xi_{in}$, and $\kappa_1\rightarrow \kappa$.  It is sufficient to model the effects of optical loss as an additional factor in the detection efficiency as detailed below.  Such a simplification is possible because loss into other optical ports or during propagation to the photodetector will only add uncorrelated noise to the light that will have zero average for the orthogonal cross correlation spectra calculated below.  With these simplifications we have,  

\begin{gather}
	x(\omega)=-\hbar G \bar{a} \chi_m(\omega) \chi_c(\omega) \sqrt{\kappa} (\xi_{in}(\omega)+\xi_{in}^{\dag}(\omega))+\chi_m(\omega)  \delta F_{th}(\omega)\\
	d(\omega)=\chi_c(\omega) (-i G \bar{a} x(\omega)+\sqrt{\kappa} \xi_{in}(\omega))
\end{gather}

	Using this solution we can compute the input and output optical noise quadrature operators.  The input amplitude and phase quadrature operators are given by,
\[
\delta X_I^{in}(\omega)=(\xi_{in}(\omega)+\xi_{in}^{\dag}(\omega)) \qquad \qquad \delta X_Q^{in}(\omega)=-i(\xi_{in}(\omega)-\xi_{in}^{\dag}(\omega)).
\]
We employ the boundary condition $a_{out}=a_{in}-\sqrt{\kappa} a$ and write $a_{out}=(\bar{a}_{out}+d_{out}(t)) e^{-i \omega_p t}$.  We then define $\delta X_I^{out}=d_{out}+d_{out}^{\dag}$ and $\delta X_Q^{out}=-i(d_{out}-d_{out}^{\dag})$.\\ 
\begin{gather}
	\delta X_I^{out}(\omega)=\delta X_I^{out,0}(\omega)\label{eq:dXIout}\\
	\delta X_Q^{out}(\omega)=\delta X_Q^{out,0}(\omega)+2\sqrt{\kappa} \chi_c(\omega) G \bar{a} \chi_m(\omega)\delta F_{th}(\omega) +2 \kappa \hbar G^2 \bar{a}^2 |\chi_c(\omega)|^2 \chi_m(\omega)\delta X_I^{out,0}(\omega) \label{eq:dXQout}
\end{gather}
The output shot noise in each quadrature in the absence of optomechanical coupling ($G\rightarrow0$) is given by,
\[
	\delta X_I^{out,0}(\omega)=\frac{\kappa/2+i\omega}{\kappa/2-i\omega} \delta X_I^{in}(\omega) \qquad \qquad \delta X_Q^{out,0}(\omega)=\frac{\kappa/2+i\omega}{\kappa/2-i\omega} \delta X_Q^{in}(\omega).
\]
We can rewrite Eq.~\ref{eq:dXQout} to match the form found in the main text,
\begin{gather}
\delta X_Q^{out}(\omega)=\delta X_Q^{out,0}(\omega)+\alpha(\omega) \delta F_{th}(\omega)+\beta(\omega) \delta X_I^{out,0}(\omega)\\
\alpha(\omega)=2\sqrt{\kappa} \chi_c(\omega) G \bar{a} \chi_m(\omega) \qquad \qquad \beta(\omega)=2 \kappa \hbar G^2 \bar{a}^2 |\chi_c(\omega)|^2 \chi_m(\omega) \nonumber
\end{gather}
We encapsulate optical loss due light absorbed by the cavity or emitted into unmeasured output ports as an effective detection efficiency $\epsilon_1=\kappa_{out}/\kappa$.  Other propagation losses between the cavity and detector are designated $\epsilon_2$, $\epsilon_3$, and $\epsilon_4$ as described above.  The output light is detected with optical heterodyne detection, providing access to all optical quadratures, but with a factor of two smaller signal to noise than is achievable for a measurement of only one optical quadrature.  This loss of signal to noise is accounted for with an effective detection efficiency $\epsilon_{het}=0.5$.  Together all of the losses add up to an overall effective detection efficiency $\epsilon=\epsilon_1 \times \epsilon_2 \times \epsilon_3 \times \epsilon_4 \times \epsilon_{het}$.  This loss modifies the detected signal,
\[
\delta X_I^{out}\rightarrow \sqrt{\epsilon} (\delta X_I^{out}) + \sqrt{1-\epsilon} (\xi_{l}(\omega)+\xi_{l}^{\dag}(\omega)) \qquad \qquad \delta X_Q^{out}\rightarrow \sqrt{\epsilon} (\delta X_Q^{out}) + \sqrt{1-\epsilon} (-i) (\xi_{l}(\omega)-\xi_{l}^{\dag}(\omega))
\]
where $\xi_l$ is the vaccum noise operator corresponding to the effective optical loss port.
 
\end{widetext}

\subsection{Correlation spectra}
Any output optical quadrature can be constructed as $\delta X_{\phi}^{out}=\cos(\phi)\delta X_I^{out}+\sin(\phi)\delta X_Q^{out}$.  We compute correlation spectra of the form $S_{\phi_1,\phi_2}(\omega)=\int_{-\infty}^{\infty} \langle \delta X^{out}_{\phi_1}(\omega') \delta X^{out}_{\phi_2}(\omega)\rangle_s d \omega'/2\pi$.  The notation $\langle A(\omega')B(\omega)\rangle_s=1/2(\langle A(\omega')B(\omega)\rangle+\langle B(\omega)A(\omega')\rangle)$ represents a symmetrized expectation value necessary to compare with the experimentally accessible (classical) photocurrent correlation spectra.  If $\phi_1=\phi_2$, this corresponds to an autocorrelation, or simply the power spectrum of optical quadrature $\phi_1$.  If $\phi_1$ and $\phi_2$ differ by $\pi/2$, this corresponds to a, generally complex valued, cross correlation spectrum.  The following expectation values for the optical vacuum noise are used to compute correlation spectra,
\begin{gather}
\langle \xi_{in,i}(\omega') \xi_{in,i}^{\dag}(\omega)\rangle=\langle\xi_{l}(\omega') \xi_{l}^{\dag}(\omega)\rangle=2 \pi \delta(\omega'+\omega)\label{eq:xiin}
\end{gather}
All other expectation values of vacuum noise operators are zero.  The spectrum of the thermal force is given by the fluctuation-dissipation theorem~\cite{Giovannetti2001},
\begin{widetext}
\[
\int_{-\infty}^{\infty}\langle \delta F_{th}(\omega') \delta F_{th}(\omega) \rangle_s d\omega'/2\pi = m \Gamma_m(\omega) \hbar \omega \coth\left(\frac{\hbar \omega}{2 k_b T}\right)
\]

Using Eq.~\ref{eq:dXIout}, \ref{eq:dXQout}, and \ref{eq:xiin}, the amplitude-phase cross correlation is,
\begin{gather}
S_{0,\frac{\pi}{2}}(\omega)=4 \epsilon \kappa g_0^2 \bar{a}^2 m \omega_m |\chi_c(\omega)|^2 \chi_m(\omega)
\end{gather}
and
\begin{gather*}
S_{\frac{\pi}{4},\frac{3\pi}{4}}(\omega)=4 \epsilon \kappa g_0^2 \bar{a}^2 m \omega_m |\chi_c(\omega)|^2 |\chi_m(\omega)|^2 m \Gamma_m(\omega) \omega \coth\left(\frac{\hbar \omega}{2 k_b T}\right)+\epsilon |\beta(\omega)|^2 +i \mathrm{Im}\{S_{0,\frac{\pi}{2}}(\omega)\}
\end{gather*}
We identify $|\chi_m(\omega)|^2 m \Gamma_m(\omega) \omega $ as $\mathrm{Im}\{\chi_m(\omega) \}$.  In the current experiment, the second term representing the RPSN driven motion is much smaller than the first term representing the thermal motion, as long as $C |\chi_c(\omega_m)|^2 \ll  k_b T/\hbar \omega_m$.  Then we have to a good approximation,
\begin{gather}
S_{\frac{\pi}{4},\frac{3\pi}{4}}(\omega) \approx 4 \epsilon \kappa g_0^2 \bar{a}^2 m \omega_m |\chi_c(\omega)|^2 \mathrm{Im}\{\chi_m(\omega)\} \coth\left(\frac{\hbar \omega}{2 k_b T}\right)+i \mathrm{Im}\{S_{0,\frac{\pi}{2}}(\omega)\} \label{eq:Spi4}
\end{gather}
as presented in the main text.  Rearranging Eq.~\ref{eq:Spi4}, we arrive at our thermometry relation,
\begin{gather}
\frac{\mathrm{Re} \{S_{\frac{\pi}{4},\frac{3\pi}{4}}(\omega)\} }{\mathrm{Im} \{S_{\frac{\pi}{4},\frac{3\pi}{4}}(\omega)\}}=\coth\left(\frac{\hbar \omega}{2 k_B T}  \right)
\end{gather}

\begin{figure*}
\includegraphics[width=.75 \textwidth]{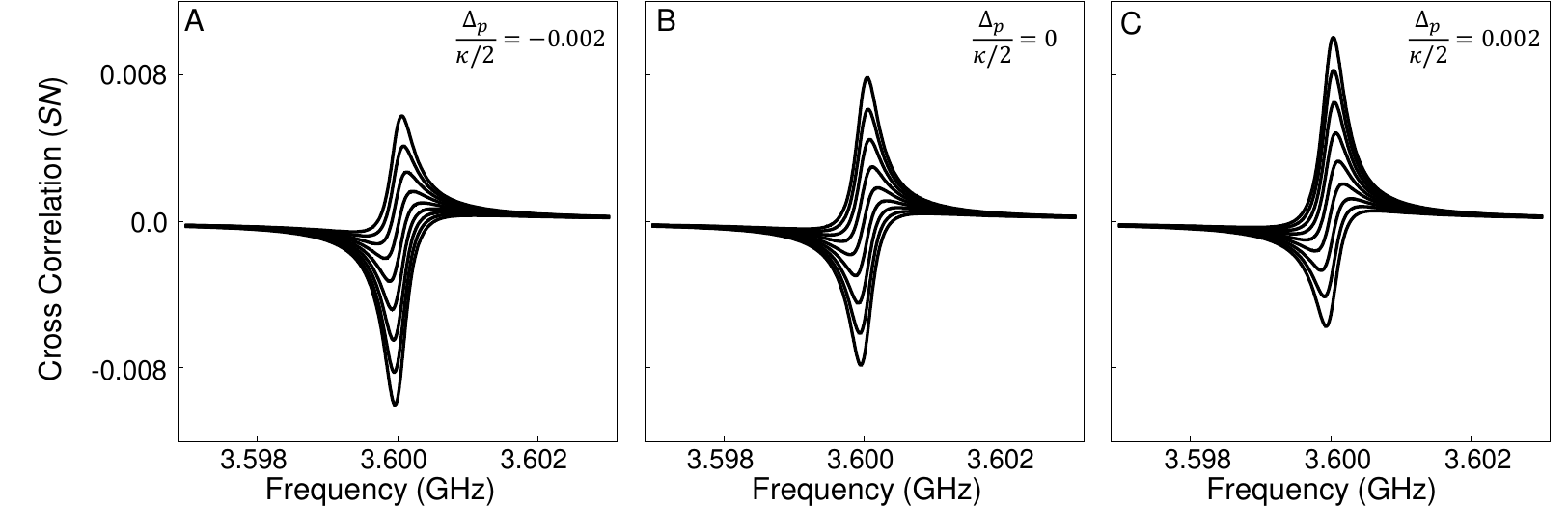}

\caption{Simulated cross correlation spectra.  $\mathrm{Re}\{S_{\phi,\phi+\frac{\pi}{2}}(\omega)\}$ are plotted for $\phi$ from -0.004 to 0.004 in steps of 0.001 from top to bottom of each panel.  The simulations use the full solution to the equations motion for $\Delta_p \neq 0$.  The probe laser detuning is varied from panel to panel according to $2 \Delta_p/\kappa=-0.002$ in (A), $2 \Delta_p/\kappa=0$ in (B), and $2 \Delta_p/\kappa=0.002$ in (C).  Parameters are chosen to approximately match the experimental conditions of the data in Fig.~2 of the main text. In each panel the correlation signal is purely dispersive for $\phi \approx 2 \Delta_p/\kappa$. }
\end{figure*}

\end{widetext}

\end{document}